\documentclass[%
reprint,
superscriptaddress,
 amsmath,amssymb,amsfonts,
 aps,
 prb,
 citeautoscript,
floatfix,
longbibliography,
]{revtex4-2}

\usepackage[utf8]{inputenc}
\usepackage[english]{babel}
\usepackage[T1]{fontenc}
\usepackage{float}
\usepackage{graphicx}% Include figure files
\usepackage{dcolumn}% Align table columns on decimal point
\usepackage{amsmath}
\usepackage{amssymb}
\usepackage{bm}% bold math
\usepackage{xcolor}
\usepackage{listings}
\usepackage{enumitem}
\usepackage{epstopdf}
\usepackage[normalem]{ulem}
\usepackage{hyperref}% add hypertext capabilities
\hypersetup{
    colorlinks=true,
    linkcolor=blue,
    citecolor=blue,
    filecolor=blue,      
    urlcolor=blue,
    }

\newcommand{\FS}[1]{\left\langle #1 \right\rangle_{\scriptscriptstyle\mathrm{FS}}}
 
\newcommand{\kb}{k_\mathrm{B}}
\newcommand{\tc}{T_\mathrm{c}}

\newcommand{\RR}{\bm{R}}
\newcommand{\RRprime}{\bm{R}^{\prime}}

\newcommand{\vF}{{\bm{v}_\mathrm{F}}}
\newcommand{\vA}{{\bm{A}}}

\newcommand{\pF}{\bm{p}_\mathrm{F}}
\newcommand{\ps}{\bm{p}_\mathrm{s}}
\newcommand{\vj}{\bm{j}}
\newcommand{\vBB}{\bm{B}}
\newcommand{\vq}{\bm{q}}
\newcommand{\vnabla}{\bm{\nabla}}
\newcommand{\vDs}{{\bm{D}}_{\rm s}}

\newcommand{\NF}{\mathcal{N}_\mathrm{F}}

\newcommand{\gammaE}{\gamma_\mathrm{E}}

\allowdisplaybreaks

\begin{document}

\title{Impurity strength--temperature phase diagram with phase crystals and competing time-reversal symmetry breaking states in nodal $d$-wave superconductors}

\author{K. M. Seja}
\email[e-mail:]{seja@chalmers.se}
\affiliation{Department of Microtechnology and Nanoscience - MC2,
Chalmers University of Technology,
SE-41296 G\"oteborg, Sweden}

\author{N. Wall-Wennerdal}
\affiliation{Department of Microtechnology and Nanoscience - MC2,
Chalmers University of Technology,
SE-41296 G\"oteborg, Sweden}

\author{T. L\"ofwander}
\affiliation{Department of Microtechnology and Nanoscience - MC2,
Chalmers University of Technology,
SE-41296 G\"oteborg, Sweden}

\author{A. M. Black-Schaffer}
\affiliation{
Department of Physics and Astronomy, Uppsala University, Box 516, SE-751 20 Uppsala, Sweden}

\author{M. Fogelstr\"om}
\affiliation{Department of Microtechnology and Nanoscience - MC2,
Chalmers University of Technology,
SE-41296 G\"oteborg, Sweden}
\affiliation{
Nordita, KTH Royal Institute of Technology and Stockholm University, Hannes Alfv\'{e}ns v\"{a}g 12, 10691 Stockholm, Sweden
}

\author{P. Holmvall}
\email[e-mail:]{patric.holmvall@physics.uu.se}
\affiliation{
Department of Physics and Astronomy, Uppsala University, Box 516, SE-751 20 Uppsala, Sweden}

\date{\today}

\begin{abstract}
Phase crystals are a class of nonuniform superconducting ground states characterized by spontaneous phase gradients of the superconducting order parameter.
These phase gradients nonlocally drive periodic currents and magnetic fields, thus breaking both time-reversal symmetry and continuous translational symmetry.
The phase crystal instability is generally triggered by negative and inhomogeneous superfluid stiffness. 
Several scenarios have been identified that can realize phase crystals, especially flat bands at specific edges of unconventional nodal superconductors.
Motivated by omnipresent disorder in all materials, we employ the ${t}$-matrix approach within the quasiclassical theory of superconductivity to study the emergence of phase crystals at edges of a nodal $d$-wave superconductor.
We quantify the full phase diagram as a function of the impurity scattering energy and the temperature, with full self-consistency in the impurity self energies, the superconducting order parameter, and the vector potential.
We find that the phase crystal survives even up to $\sim 40\text{--}50\%$ of the superconducting critical impurity strength in both the Born and unitary scattering limits. 
Finally, we show how mesoscopic finite-size effects induce a competition with a state still breaking time-reversal symmetry but with translationally invariant edge currents. 

\end{abstract}

\maketitle

\section{Introduction}

Superconductivity is characterized by the phase $\chi$ of the electron pair condensate, which breaks the $U(1)$-symmetry, and quantified by the superconducting order parameter $\Delta = |\Delta|e^{i\chi}$.
Gradients in this phase generate a superflow, %$\ps(\RR) = (\hbar/2)\vnabla\chi(\RR) - (e/c)\vA(\RR)$
which couples to the kinetic energy via the superfluid stiffness, %$F_{\rm kin} = \int\int d\RR d\RRprime \ps(\RR)\cdot\vDs(\RR,\RRprime)\cdot\ps(\RRprime) + \mathcal{O}(\ps^4)$, with 
%center-of-mass coordinate $\RR$,
%reduced Planck constant $\hbar$ %elementary charge $e$, speed of light $c$, vector potential $\vA$
%and superfluid stiffness $\vDs$.
where a positive stiffness %$\vDs(\RR,\RRprime)>0$
yields a uniform and rigid phase in order to minimize the kinetic energy. % $\vnabla\chi = 0$ (to minimize $F_{\rm kin}$).
However, external fields can favor nonuniform superconducting states such as the Abrikosov-vortex state~\cite{Abrikosov1957,Tinkham2004} or the \textit{amplitude} instabilities of Fulde-Ferrell $\Delta(\RR)\propto\Delta_{\vq}e^{i\vq\cdot\RR}$~\cite{Fulde1964Aug} or Larkin-Ovchinnikov $\Delta(\RR) \propto \Delta_{\vq}\cos(\vq\cdot\RR)$~\cite{Larkin1964Sep} type.

Recently, it has been shown~\cite{Hakansson2015Sep} that even in the absence of external fields, and deep below the superconducting transition temperature $T_{\mathrm{c}}$ where $\Delta$ is already well-established, there can exist %negative superfluid-density correlations $\vDs(\RR,\RRprime)<0$ triggering
another nonuniform superconducting state. This state occurs at a 
spontaneous phase transition at $T^* \sim 0.2\tc$ and is a \textit{phase} instability $\chi(\RR) \propto C_{\vq}\cos(\vq\cdot\RR)$, referred to as a ``phase crystal''~\cite{Holmvall2020Jan}.
Here, $\vq$ is the instability wave vector of this nonuniform, but still equilibrium, ground state.
Phase crystals are thus an emergent superconducting phenomenon, a priori not requiring any additional interactions or fields besides superconductivity itself.
Specifically, it is a negative superfluid stiffness that yields the spontaneous phase gradients and furthermore drives nonlocal equilibrium currents and magnetic fields.
The phase crystal state is thus associated with both time-reversal symmetry breaking (TRSB) and continuous translational symmetry breaking (TSB)~\cite{Holmvall2018Jun}.

\begin{figure}[t!]
	\includegraphics[width=\columnwidth]{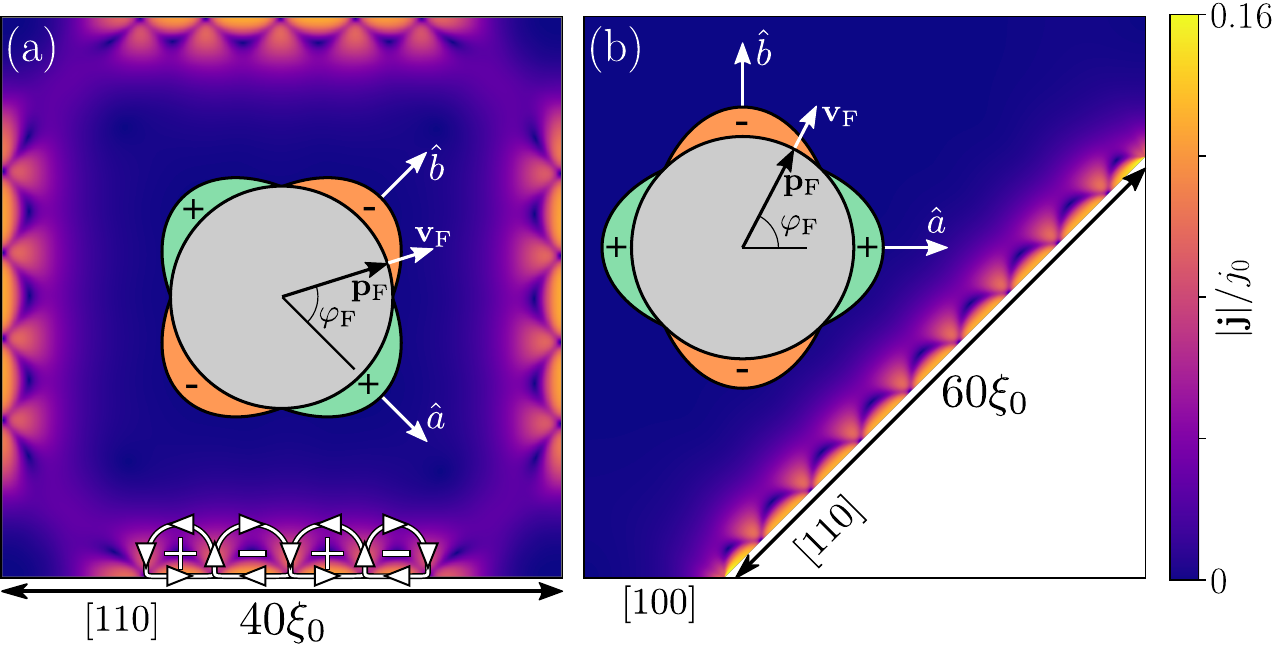}
    \caption{Sketch of the two different $d_{x^2-y^2}$-wave superconductor geometries considered in this work.
    (a) Square grain with edges rotated $45^{\circ}$ degrees with respect to the crystal $ab$-axes, each of which is pair-breaking due to resonant Andreev reflection.
    (b) Grain with only one pair-breaking edge, while all other edges are aligned with the $ab$-axes.
   Heatmaps show the magnitude of the spontaneous loop currents [arrows at bottom edge in (a)] produced by the phase crystal instability and computed self-consistently at temperature $T=0.11\tc$ in the absence of impurities. Inset illustrates the nodal $d_{x^2-y^2}$-wave gap symmetry (colors) on the Fermi surface (gray), with vectors indicating the Fermi velocity $\vF$ at Fermi momentum $\pF$ and angle $\varphi_{\rm F}$ relative to the crystal $ab$-axes. Here, $j_0$ is a natural unit (see Sec.~\ref{sec:theory}).}
    \label{fig:systemSketch}
\end{figure}

Phase crystals have been predicted to emerge e.g.~due to flat bands at zero energy associated with Andreev bound states (ABS) appearing at pair-breaking edges of unconventional nodal superconductors~\cite{Hakansson2015Sep}, but also in conventional superconductor-ferromagnet heterostructures~\cite{Holmvall2020Jan}.
In this work we consider edges of nodal $d_{x^2-y^2}$-wave superconductors which are fully misaligned with respect to the crystal $ab$-axes, e.g.~$[110]$ edges as visualized in Fig.~\ref{fig:systemSketch}, as they represent an easily accessible experimental system.
Here, quasiparticles accumulate a $\pi$-phase factor as they scatter between order parameter lobes of different sign, consequently inducing resonant Andreev reflection that break superconducting pairs and form flat band ABS at zero energy~\cite{Hu1994,lofwander_andreev_2001}. Such zero-energy ABS have been experimentally observed as a zero-bias conductance peak~\cite{Becherer1993Jun,Kashiwaya1994Dec,Kashiwaya1995Jan}.
These ABS are further topological~\cite{Ryu2002Jul,Sato2011Jun,Ikegaya2017Jun,Nagai2017Aug} but at the same time thermodynamically unstable due to their large ground-state degeneracy, where any shift to finite energy has been shown to lower the free energy~\cite{Vorontsov2018}.
Such an energy shift is consistent with an experimentally observed temperature-independent broadening of the flat bands~\cite{Geerk1988Sep,Lesueur1992Feb,Covington1997Jul}.
Several mechanisms, in additional to phase crystal formation, have been proposed to explain this spontaneous broadening, including a subdominant superconducting order~\cite{Matsumoto1995Mar,Fogelstrom1997Jul,Sigrist1998Jun}, ferromagnetism~\cite{Honerkamp2000May,Potter2014Mar}, and spontaneous currents~\cite{Higashitani1997Apr,Barash2000Sep,Lofwander2000Dec}.
Notably, all these scenarios are related to spontaneous TRSB and leads to currents and net magnetic signatures observable with scanning probes~\cite{Persky:2022}.
However, experimental detection remains controversial~\cite{Carmi2000Apr,Tsuei2000Oct,Neils2002Jan,Kirtley2006Feb,Saadaoui2011Feb}.
Phase crystals can resolve this controversy~\cite{Hakansson2015Sep,Chakraborty2022Apr}; it is typically the most energetically favorable scenario~\cite{Hakansson2015Sep,Seja2024Aug}, has zero net magnetic signature beyond the superconducting coherence length $\xi_0$ and is thus consistent with previous experiments~\cite{Holmvall2018Jun,Wennerdal2020Nov}, and results in an ABS peak broadening fully consistent with experiments~\cite{Chakraborty2022Apr}.

Multiple different theoretical frameworks have been used to demonstrate the spontaneous appearance of phase crystals including for nodal superconductors, from quasiclassical theory~\cite{Hakansson2015Sep,Holmvall2018Jun,Holmvall2020Jan,Seja2022Oct}, phenomenological Ginzburg-Landau theory~\cite{Holmvall2020Jan}, to fully microscopic theories~\cite{Wennerdal2020Nov}, also including strong electron-electron correlations as appropriate for the high-temperature cuprate superconductors~\cite{Chakraborty2022Apr}.
Furthermore, phase crystals have been studied in the presence of external magnetic fields~\cite{Holmvall2018Jun,Holmvall2023Mar}, various geometric~\cite{Holmvall2018Mar,Holmvall2019May} and Fermi-surface effects~\cite{Wennerdal2020Nov}, competing superconducting~\cite{Hakansson2015Sep} and magnetic orders~\cite{Seja2024Aug}, topological superconductivity~\cite{Bonetti2024May}, and nonmagnetic impurities~\cite{Chakraborty2022Apr,Seja2022Oct}.
In particular, nonmagnetic impurities are omnipresent in all materials and phase crystals have so far only been predicted to be robust against such impurities when including the additional effects of strong electron-electron interactions~\cite{Chakraborty2022Apr}.

In order to aid experimental efforts to realize and detect phase crystals, we here set out to explicitly quantify the robustness of phase crystals in the presence of impurities also in the weak coupling regime. We achieve this by quantifying the full ground-state phase diagram as a function of temperature $T$ and impurity scattering energy $\Gamma$,
considering general scattering phase shifts $\delta_0$ between the Born ($\delta_0=0$) and the unitary ($|\delta_0|=\pi/2$) scattering limits, with full self-consistency simultaneously in the impurity self-energies $h_{\rm imp}$, superconducting order parameter $\Delta$ and vector potential $\vA$.
We accomplish this by using the well-established $t$-matrix approach~\cite{Buchholtz1979Jun,Buchholtz1981Jun,Xu1995Jun} within the quasiclassical theory of superconductivity~\cite{eilenberger_transformation_1968,Larkin1969, Eliashberg1971}.

We begin our study by establishing how nonmagnetic impurities by themselves broaden the surface ABS in nodal $d_{x^2-y^2}$-wave superconductors as a function of both $\Gamma$ and $\delta_0$, finding the strongest (weakest) broadening in the Born (unitary) scattering limit, which is consistent with analytic calculations~\cite{Poenicke1999,Zare2008Sep}.
We then take these results and study the influence of nonmagnetic impurities on the phase crystal and its emergent current loops, the latter also visualized in Fig.~\ref{fig:systemSketch}.
Since ABS at different pair-breaking edges may also hybridize in mesoscopic systems~\cite{Iniotakis2005Jun,Sauls2011Dec,Wu2023Sep}, we choose to study systems whose geometries host either multiple or single pair-breaking edges, i.e.~where the hybridization between pair-breaking edges is either significant as in Fig.~\ref{fig:systemSketch}(a) or completely absent as in Fig.~\ref{fig:systemSketch}(b).
We find find that the ABS impurity-broadening reduce both the magnitude and number of current loops, and that these effects can be further enhanced by edge-edge hybridization.
In particular, nonmagnetic impurities cause the effective system size $\mathcal{D}/\xi(\Gamma)$ to shrink due to the coherence length $\xi(\Gamma)$ increasing with $\Gamma$.
In fact, for sufficiently small $\mathcal{D}/\xi(\Gamma)$, we find that the hybridization can become so strong that the phase crystal evolves into another competing TRSB phase which has translationally invariant edge currents, originally proposed by Vorontsov~\cite{Vorontsov2009Apr} but then in a clean system within a slab geometry.
%($\vq\to0$)~\cite{Vorontsov2009Apr}.
We directly quantify the size effects and the competition between different states within the full ground-state phase diagram, where we compute both the TRSB transition temperature $T^*$ and the superconducting critical temperature $\tc$ as functions of $\Gamma$, for different scattering phase shifts $\delta_0$ and the two geometries in Fig.~\ref{fig:systemSketch}.
In particular, we find that the phase crystal is remarkably robust, surviving up until $\sim40\text{--}50\%$ of the bulk critical impurity strength, from the Born to the unitary scattering limits, respectively.
Furthermore, we find that the superconducting transition temperature $\tc$ can be strongly suppressed by finite-size effects in the presence of impurities, while the transition temperature $T^*$ instead increases by these finite-size effects.
Our results thus imply that phase crystals should be robust enough to be present under experimental conditions in generic nodal $d$-wave superconductors.

The rest of the work is outlined as follows.
In Sec.~\ref{sec:theory}, we present our theoretical formalism.
In Sec.~\ref{sec:ABS} we study the impurity broadening of the surface ABS.
In Sec.~\ref{sec:phase_crystal} we study the influence of impurities on the phase crystal.
In Sec.~\ref{sec:phase_diagram} we present the ground-state phase diagram.
In Sec.~\ref{sec:summary}, we summarize our work and give an outlook of interesting future research directions.

\section{Theory and Modeling}
\label{sec:theory}
In this section we introduce the theory and modeling used in this work.

\subsection{Quasiclassical theory of superconductivity}
\label{sec:theory:quasiclassics}
To study phase crystals and impurities, we use the quasiclassical theory of superconductivity~\cite{eilenberger_transformation_1968,Larkin1969, Eliashberg1971}, solving the time-independent and equilibrium form of the Eilenberger equation~\cite{eilenberger_transformation_1968}
\begin{align}
i \hbar &\vF \cdot \nabla \hat{g}^\mathrm{M}(\RR,\pF,\varepsilon_n) 
\nonumber
\\
&+ \bigl[i \varepsilon_n \hat{\tau}_3  - \hat{h}^\mathrm{M}(\RR,\pF,\varepsilon_n), \hat{g}^\mathrm{M}(\RR,\pF,\varepsilon_n) \bigr] = 0,
\label{eq:transportequation}
\end{align}
for the quasiclassical Green's function $\hat{g}^\mathrm{M}$ that depends on spatial coordinate $\RR$, the momentum direction on the Fermi surface $\pF$, and the Matsubara (M) frequency $\varepsilon_n$.
Here, $\hbar$ is the reduced Planck constant.
Assuming a circular Fermi surface, all momenta $\pF$ can be parametrized in terms of a single angle $\varphi_\mathrm{F} \in [0, 2\pi)$.
Additionally, Eq.~\eqref{eq:transportequation} contains the Fermi velocity $\mathbf{v}_\mathrm{F}(\varphi_{\rm F})$ and the self-energy matrix $\hat{h}^\mathrm{M}$.
By $[A,B]$ we denote the commutator between matrices $A$ and $B$, the hat-symbol "$\hat{~}$" indicates a matrix in particle-hole (Nambu) space, while $\hat{\tau}_3$ is the third Pauli matrix. In addition to satisfying Eq.~\eqref{eq:transportequation}, the Green's function $\hat{g}^\mathrm{M}$ also has to obey the normalization condition~\cite{eilenberger_transformation_1968}
\begin{equation}
\hat{g}^\mathrm{M}(\RR,\pF,\varepsilon) \hat{g}^\mathrm{M}(\RR,\pF,\varepsilon) = -\pi^2.
\label{eq:normalizationCondition}
\end{equation}
As a $2 \times 2$ matrix in particle-hole space, the Green's function $\hat{g}^\mathrm{M}$ can be written as 
\begin{align}
\hat{g}^\mathrm{M}
=
\begin{pmatrix}
g^\mathrm{M} & f^\mathrm{M}
\\
\tilde{f}^\mathrm{M} & \tilde{g}^\mathrm{M}
\end{pmatrix}.
\label{eq:GreensFunctionElements}
\end{align}
The diagonal elements are the single-quasiparticle Green's function, while the off-diagonal elements describe anomalous superconducting correlations.
Quantities with and without a tilde are related via particle-hole conjugation,
\begin{align}
\tilde{A}(\RR,\pF,\varepsilon) = A^*(\RR,-\pF,-\varepsilon^*),
\label{eq:TildeSymmetry}
\end{align}
For clarity, we assume spin-degeneracy in this work such that all four elements of $\hat{g}^\mathrm{M}$, which in general are $2\times2$-spin matrices, reduce to scalar functions. 
The matrix elements in Eq.~\eqref{eq:GreensFunctionElements} can thus be parametrized in terms of two scalar coherence functions $\gamma$ and $\tilde{\gamma}$, the so-called Riccati parametrization~\cite{Nagato1993, Schopohl1995, Schopohl1998}. This parametrization greatly simplifies the numerical solution of Eq.~\eqref{eq:transportequation}.

The self-energy matrix $\hat{h}^\mathrm{M}$ is also a $2\times2$ matrix in Nambu space.
Here, the self-energy captures all the interactions in the system, which in our work consists of a mean-field superconducting order parameter (MF), scalar impurity scattering (imp), and electromagnetic (EM) interactions
\begin{align}
    \hat{h}^\mathrm{M} = \hat{h}^\mathrm{M}_\mathrm{MF} + \hat{h}^\mathrm{M}_\mathrm{imp} + \hat{h}^\mathrm{M}_\mathrm{EM}.
    \label{eq:selfenergy_contributions}
\end{align}
We start by describing $\hat{h}^\mathrm{M}_\mathrm{MF}$, while $\hat{h}^\mathrm{M}_\mathrm{imp}$ and $\hat{h}^\mathrm{M}_\mathrm{EM}$ are described in Sec.~\ref{sec:theory:impurities} and Sec.~\ref{sec:theory:electromagnetics}, respectively.

We are interested in an order parameter $\Delta(\RR,\pF)$ with spin-singlet $d_{x^2-y^2}$-wave pairing symmetry,
\begin{align}
    \label{eq:superconducting_self_energy}
    \hat{h}^\mathrm{M}_\mathrm{MF}
= 
\begin{pmatrix}
0 & \Delta \\
-\tilde{\Delta} & 0 
\end{pmatrix},
\end{align}
with scalar matrix elements
\begin{align}
\Delta(\RR,\pF) = \Delta(\RR) \eta_d(\varphi_F) =
\Delta(\RR) \cos \left( 2 \varphi_\mathrm{F}- 2 \alpha \right). 
\end{align}
Here, $\alpha$ is the misalignment angle between surface normals and the crystal axis, and $\Delta(\RR)$ is a (generally complex) scalar number.
The scalar value $\Delta(\RR)$ is determined from the self-consistency equation
\begin{align}
    \label{eq:gap_equation}
    \Delta(\RR) = \lambda_d \mathcal{N}_\mathrm{F} \kb T \sum\limits_{|\varepsilon_n| < \varepsilon_{\rm c}}
\FS{ \eta_d(\pF) f^\mathrm{M}(\RR, \varepsilon_n, \pF) }, 
\end{align}
with $d$-wave coupling constant $\lambda_d$, normal-state density of states $\mathcal{N}_\mathrm{F}$ (per spin), Boltzmann constant $\kb$, temperature $T$, Matsubara sum cutoff $\varepsilon_{\rm c}$, and the Fermi-surface average~\cite{Graf1993May}
\begin{equation}
\FS{\dots} = \int_0^{2\pi}\frac{d\varphi_\mathrm{F}}{2\pi}(\dots).
\label{eq:FermiSurfaceAverage}
\end{equation}
We eliminate the coupling constant $\lambda_d$ and cutoff $\varepsilon_{\rm c}$ in favor of the critical temperature $T_{\mathrm{c}}$~\cite{Grein2013Aug}.

\subsection{Clean systems}
\label{sec:theory:clean_system}

In a uniform environment, such as bulk or at an edge aligned with the order parameter lobes (e.g.~a $[100]$ edge for $d_{x^2-y^2}$-wave), the spin-degenerate propagator takes the bulk form
\begin{align}
\hat{g}^{\rm M,R}_{\mathrm{bulk}}(\pF,\varepsilon)
=
\frac{\pi}{\Lambda(\varepsilon)}\begin{pmatrix}
-\varepsilon & \Delta(\pF)
\\
-\tilde{\Delta}(\pF) & \varepsilon^*
\end{pmatrix},
\label{eq:GreensFunctionBulk}
\end{align}
here in $2\times2$ particle-hole space where $\Lambda(\varepsilon) \equiv \sqrt{|\Delta|^2 - \varepsilon^2}$.
Using the clean-limit critical temperature $T_{\mathrm{c},0}$ we define the bulk superconducting coherence length $\xi_0 \equiv \hbar v_\mathrm{F}/(2 \pi \kb T_\mathrm{c,0})$
%\begin{align}
%\xi_0 \equiv \frac{\hbar v_\mathrm{F}}{2 \pi \kb T_\mathrm{c,0}},
%\label{eq:xi0Definition}
%\end{align}
which is the characteristic length scale of superconductivity.

Assuming a uniform order parameter amplitude $\Delta_0$, the propagators can also be derived for other interface orientations, qualitatively capturing interesting surface effects~\cite{Sauls2011Dec}.
For instance, in a semi-infinite system with an interface aligning with the order parameter nodes, i.e.~a maximally pair-breaking interface such as $[110]$ (as in Fig.~\ref{fig:systemSketch}), resonant ABS~\cite{Buchholtz1981Jun,Hu1994,lofwander_andreev_2001} yields an additional surface term in the propagator~\cite{Holmvall2019Nov}
\begin{align}
\nonumber
\hat{g}^{\rm M,R}(\RR,\pF,\varepsilon)
& = \hat{g}^{\rm M,R}_{\mathrm{bulk}}(\pF,\varepsilon)\left(1 - e^{-y/\xi_y}\right) \\
\label{eq:GreensFunctionSurface:1}
& + \hat{g}^{\rm M,R}_{\mathrm{surface}}(\pF,\varepsilon)e^{-y/\xi_y},\\
\label{eq:GreensFunctionSurface:2}
\hat{g}^{\rm M,R}_{\mathrm{surface}}(\pF,\varepsilon) & = \frac{\pi}{\varepsilon}\begin{pmatrix}
\Lambda(\varepsilon) & -is\Delta(\pF)
\\
s\tilde{\Delta}(\pF) & -\Lambda(\varepsilon)
\end{pmatrix},
\end{align}
with surface normal $\hat{y}$, effective coherence length $\xi_y \equiv \hbar |v_{\mathrm{F},y}|/2\Lambda(\varepsilon)$, where $v_{\mathrm{F},y} \equiv v_{\rm F}\sin(\varphi_{\rm F})$ and  $s\equiv\operatorname{sgn}(v_{\mathrm{F},y})$.
We note that for the equilibrium Matsubara propagator ($\varepsilon \to i\varepsilon_n$), the pre-factor in Eq.~(\ref{eq:GreensFunctionSurface:2}) is inversely proportional to the temperature, $\varepsilon_n^{-1} \propto T^{-1}$, thus highly important at low temperature.

\subsection{Impurity self-energies}
\label{sec:theory:impurities}
nonmagnetic impurities are introduced into our approach through the self-energy $\hat{h}^\mathrm{M}_\mathrm{imp}$ in Eq.~(\ref{eq:selfenergy_contributions}).
Specifically, we use the well-established $t$-matrix approach~\cite{Buchholtz1979Jun,Buchholtz1981Jun}, which importantly includes a proper treatment of the averaging over impurity realizations~\cite{Rammer:2007}.
The impurity self-energy in Eq.~\eqref{eq:selfenergy_contributions} generally has both diagonal and off-diagonal elements
\begin{equation}
\label{eq:impurity_matrix}
\hat{h}^\mathrm{M}_\mathrm{imp} = 
\begin{pmatrix}
\Sigma_\mathrm{imp} & \Delta_\mathrm{imp}
\\
-\tilde{\Delta}_\mathrm{imp} & \tilde{\Sigma}_\mathrm{imp}
\end{pmatrix}.
\end{equation}
We consider an impurity self-energy that describes momentum-isotropic ($s$-wave) scattering on a dilute impurity concentration $n_i$ with scattering potential $u_0$, and with an averaging over impurity position.
Under these assumptions, the $t$-matrix equation for the Matsubara self-energy
\begin{align}
    \label{eq:impurity_self_energy}
    \hat{h}^\mathrm{M}_\mathrm{imp} = n_\mathrm{i} \hat{t}^\mathrm{M},
\end{align}
can be solved in the noncrossing approximation yielding~\cite{Rammer:2007}
\begin{align}
    \label{eq:impurity_t_matrix}
    \hat{t}^\mathrm{M} &= \frac{u_0 \hat{1} + u_0^2 \NF \FS{\hat{g}^\mathrm{M}}}{\hat{1} - \left[  u_0 \NF \FS{\hat{g}^\mathrm{M} }\right]^2   }.
\end{align}
As a re-parametrization of the two scattering model parameters $n_\mathrm{i}$ and $u_0$,
we use the scattering energy $\Gamma_u$ and scattering phase shift $\delta_0$~\cite{Xu1995Jun}  
\begin{align}
\Gamma_u &\equiv \frac{n_\mathrm{i}}{\pi \NF},
\\
\delta_0 &\equiv \arctan (\pi u_0 \NF).
\end{align} 
For a $d$-wave order parameter, scalar impurity scattering is pair-breaking and suppresses the order parameter~\cite{Buchholtz1981Jun,Xu1995Jun}.
The pair-breaking strength can be defined in terms of the energy
\begin{align}
\Gamma \equiv \Gamma_u \sin^2 \delta_0,
\label{eq:PairBreakingParameter}
\end{align}
where the superconducting order parameter vanishes at a critical impurity scattering energy $\Gamma_{\rm c}(T)$, or conversely at a critical temperature $T_{\mathrm{c}}(\Gamma)$.
We compute these quantities fully self-consistently, and compare our results with the corresponding expressions obtained via the well-known Abrikosov-Gorkov formula~\cite{Abrikosov1960Dec, Xu1995Jun} for a bulk $d$-wave superconductor
\begin{align}
\ln \frac{ T_{\mathrm{c},0} }{ T_{\mathrm{c}} } 
= \psi \left( \frac{1}{2} + \frac{\Gamma}{2\pi \kb T_{\mathrm{c}}} \right) 
- 
\psi \left( \frac{1}{2} \right),
\label{eq:AbrikosovGorkovFormula}
\end{align}
with the digamma function $\psi(x)$. Introducing the bulk definitions $T_{\rm c,0}\equiv \tc(\Gamma=0)$ and $\Gamma_{\rm c,0} \equiv \Gamma(T=0)$, we obtain
$\Gamma_{\rm c, 0}/(2\pi\kb T_{\mathrm{c},0}) = e^{-\gammaE}/4 \approx 0.14$, where $\gammaE$ is the Euler-Mascheroni constant.
For comparison, the bulk $d$-wave gap at $T \to 0$ and in the clean limit is $\Delta_0/(2\pi\kb T_{\mathrm{c},0}) = e^{-\gammaE - 1/2}/\sqrt{2} \approx 0.24$ (the corresponding maximal gap in the DOS is $\sqrt{2}$ times this value).

In addition to self-consistently solving the impurity self-energies from Eqs.~(\ref{eq:impurity_matrix})--(\ref{eq:impurity_t_matrix}) with general scattering phase shifts $\delta_0$, we in this work also focus on two important limiting cases for the scattering self-energy.
First, the Born scattering limit of weak impurity scattering, described by $\delta_0 \rightarrow 0$ and $\Gamma_u \rightarrow \infty$ while $\Gamma$ is kept constant.
In this case, the impurity self-energy is simply
\begin{align}
    \hat{h}^\mathrm{M}_\mathrm{Born} = \frac{\Gamma}{\pi} \FS{\hat{g}^\mathrm{M}}.
\end{align}
Second, we consider the unitary scattering limit of strong scattering where $|\delta_0| \rightarrow \pi/2$ such that $\Gamma \to \Gamma_u$. Then, the impurity self energy reads
\begin{align}
    \hat{h}^\mathrm{M}_\mathrm{uni} = -\pi \Gamma_u \frac{\FS{\hat{g}^\mathrm{M}}}{\FS{\hat{g}^\mathrm{M}}^2 }.
\end{align}
%One particular feature of this unitary scattering is the appearance of a band of impurity states at the Fermi energy.
We note that the influence of impurities on ABS has previously been studied in both the Born and unitary scattering limits, where the impurities were shown to broaden the ABS~\cite{Poenicke1999}.
Generally speaking, Born-limit scattering leads to a larger broadening of the surface states with impurity concentration, which we reproduce for finite-sized systems in Sec.~\ref{sec:ABS}.
In the above two cases impurity scattering does not induce any electron-hole asymmetry.
However, for any value of the scattering phase shift $\delta_0$ between these two limits there is indeed such an asymmetry which alters the physics~\cite{monien87a,arf88,arf89,sal96,Lofwander2004Jul,Seja2022Mar}.

\subsection{Electromagnetic interactions}
\label{sec:theory:electromagnetics}
We are interested in studying TRSB phases, usually hosting spontaneous charge-current density $\vj(\RR)$, which we compute via the Matsubara pole expansion
\begin{align}
    \label{eq:current_density}
    \mathbf{j}(\RR) \equiv 2 \kb T \mathcal{N}_\mathrm{F} |e| \sum\limits_{\varepsilon_n > 0} \FS{\vF g^\mathrm{M}(\RR, \varepsilon_n, \pF)},
\end{align}
which we quantify in natural units $j_0 \equiv \hbar|e|v_{\rm F}^2N_{\rm F}/\xi_0$, with elementary charge $e=-|e|$. 
The charge-current density induces a magnetic flux density $\vBB(\RR)$ and vector potential $\vA(\RR)$ through Amp{\`e}re's law 
\begin{align}
    \label{eq:ampere_law}
    \vnabla\times\vBB(\RR) = \vnabla \times \vnabla \times \vA(\RR) = \frac{4\pi}{c}\vj(\RR),
\end{align}
with speed of light $c$, where $\vA$ couples to the momentum of the quasiparticles through the self-energy term
\begin{align}
    \label{eq:selfenergy_contributions:EM}
    \hat{h}^\mathrm{M}_\mathrm{EM} = -\frac{e}{c}\vF(\pF)\cdot\vA(\RR)\hat{\tau}_3.
\end{align}
We assume the Coulomb gauge such that Eq.~(\ref{eq:ampere_law}) reduces to Poisson's equation $-\vnabla^2\vA(\RR)=(4\pi/c)\vj(\RR)$, where we use standard Green's function techniques to self-consistently solve $\vA(\RR)$ generated by the currents in Eq.~(\ref{eq:current_density}), see Ref.~\cite{Holmvall2023Mar} for further details.
We note that the magnetic flux density $\vBB(\RR)$ has been studied in detail for the phase crystal state before~\cite{Hakansson2015Sep,Holmvall2017,Holmvall2018Mar,Holmvall2018Jun}.
%As a quantitative measure for the total current in the system, we use the area-integrated absolute value of the charge-current density
%\begin{align}
%   \mathcal{J} \equiv \frac{1}{\mathcal{A}}\int\limits_{\mathcal{A}} d\RR ~|\mathbf{j}(\RR)|,
%   \label{eq:AverageCurrentDefinition}
%\end{align}
%normalized by the area $\mathcal{A}$ of the system under consideration. 

\subsection{Density of states and free energy}
In order to analyze our self-consistent solutions, we compute the local density of states (LDOS)
\begin{equation}
N(\RR,\varepsilon) = 2\NF\FS{\mathcal{N}(\pF, \RR, \varepsilon)},
\label{eq:DOS_full}
\end{equation}
where
\begin{equation}
\mathcal{N}(\pF, \RR, \varepsilon) = -\frac{1}{4\pi}\mathrm{Im}~\mathrm{Tr}\left[\hat{\tau}_3 \hat{g}^{\mathrm{R}}(\pF, \RR, \varepsilon)\right]
\label{eq:DOS_angle}
\end{equation}
is the momentum-resolved density of states.
Note that this requires the retarded (R) Green's function $\hat{g}^{\mathrm{R}}$ which is obtained by solving Eq.~\eqref{eq:transportequation} with the replacement $i \varepsilon_n \rightarrow \varepsilon + i \eta$ where $\eta$ is a small positive number.

In order to obtain the ground state phase diagram, we compute the free energy of different self-consistent superconducting $(S)$ solutions with respect to the normal state $(N)$, $\Omega \equiv \Omega_{\rm S} - \Omega_{\rm N}$.  To this end, we use the functional
\begin{align}
    \Omega 
    \equiv 
    \mathcal{N}_\mathrm{F} 
    \int\!\mathrm{d}\RR~
    \Biggl\{ 
    &|\Delta(\pF, \RR)|^2 \Bigl( \ln \frac{T}{T_\mathrm{c,0}} + \pi \kb T \sum\limits_{n} \frac{1}{|\varepsilon_n|} \Bigr)
    \nonumber
    \\
    &- \pi \kb T  \sum\limits_{n}  \FS{\mathcal{I}(\pF, \varepsilon_n, \RR)}
    \Biggr\},
    \label{eq:FreeEnergyEilenberger}
\end{align}
with the functional kernel
\begin{align}
    \mathcal{I}(\pF, \varepsilon_n, \RR) \equiv 
    \frac{ \tilde{\Delta}(\pF) f^{\rm M}(\pF, \varepsilon_n) + \Delta(\pF) \tilde{f}^{\rm M}(\pF, \varepsilon_n) }{\pi + i g^{\rm M}(\pF, \varepsilon_n)}.
\end{align}
In the latter expression it is implied that all quantities on the right-hand side depend on the spatial position.
Here, Eq.~\eqref{eq:FreeEnergyEilenberger} is the Eilenberger form of the free energy~\cite{eilenberger_transformation_1968}, which has been shown to be applicable to the scenario considered in our work.
Specifically, the impurity self-energy in the $t$-matrix approximation vanishes in the more general Luttinger-Ward functional~\cite{Luttinger1960Jun, Serene1983Dec,Vorontsov2003Aug, Ali2011Feb,Virtanen2020Mar}, which reduces to the above Eilenberger form in the limit considered here~\cite{Virtanen2020Mar,Holmvall2019Nov}. 

\subsection{Model and numerical details}
\label{sec:theory:model_numerics}

For clarity, we in this work assume equilibrium, spin degeneracy, weak coupling superconductivity,
a circular Fermi surface~\cite{Graf1993May} and specular reflection boundary conditions at any superconductor-vacuum interfaces.
The influence of noncircular Fermi surfaces~\cite{Wennerdal2020Nov} or mesoscopic surface roughness~\cite{Holmvall2019May} have previously been investigated on phase crystals.
Furthermore, we solve for full self-consistency in the vector potential via finite penetration depth $\lambda_{0}$~\footnote{Using realistic values for the penetration depth $\lambda_{0} \sim 80\text{--}120\xi_0$~\cite{Walter1998Apr,Kamal1998Oct,Sonier2000Jul,Pereg-Barnea2004May}, we solve for self-consistency simultaneously in the superconducting order parameter $\Delta$, impurity self energies $h_{\rm imp}$ and gauge field $\mathbf{A}$. We find negligible influence of the self-screening by finite $\lambda_0$ at both higher and lower temperatures $T=0.12T_{\mathrm{c},0}$ and $T=0.05T_{\mathrm{c},0}$, higher and lower scattering energies $\Gamma=0.02\times2\pi\kb T_{\mathrm{c},0}$ and $\Gamma=0.002\times2\pi\kb T_{\mathrm{c},0}$, in both the Born and unitary scattering limits.}, but find negligible influence on the phase crystal and therefore focus on extreme type-II superconductivity for simplicity, via $\lambda_0 \to \infty$.

For our numerical calculations, we use the open-source software \textsc{SuperConga} that enables studies on a wide variety of phenomena using the quasiclassical theory of superconductivity~\cite{Holmvall2023Mar}.
Specifically, \textsc{SuperConga} uses the numerically stable Riccati formalism~\cite{Nagato1993,Schopohl1995,Schopohl1998} to solve the Eilenberger equation (\ref{eq:transportequation}) along ballistic trajectories defined by the Fermi velocity $\vF$, with full self-consistency simultaneously in the order parameter $\Delta$ via Eqs.~(\ref{eq:superconducting_self_energy})--(\ref{eq:gap_equation}), vector potential $\vA$ via Eqs.~(\ref{eq:current_density})--(\ref{eq:selfenergy_contributions:EM}), and impurity self-energies via Eqs.~(\ref{eq:impurity_matrix})--(\ref{eq:impurity_t_matrix}).
With appropriate initial guesses for all self-energies, this is done iteratively until the global relative error fulfills $\varepsilon_{\rm G} = \left\| O_i - O_{i-1} \right\| / \left\| O_{i-1} \right\| < \varepsilon_{\rm tol}$ at iteration $i$ in each relevant observable $O \in \{\Delta,\Omega,\mathbf{j},\ldots\}$, where $\varepsilon_{\rm tol} = 10^{-7}$ is our tolerance for self-consistency.
See Ref.~\cite{Holmvall2023Mar} for further details.
We find it sufficient to use a spatial resolution of $20$ discrete points per coherence length, an energy cutoff $\varepsilon_{\rm c} = 16 \times 2\pi\kb T_{\mathrm{c},0}$, and $250$ points to discretize the Fermi-surface averages (except when calculating the LDOS for which we use $720$ points).
We use an energy broadening $\eta = 0.005 \times 2\pi\kb T_{\mathrm{c},0}$ when computing the LDOS.

\section{Impurity broadening of ABS}
\label{sec:ABS}

\begin{figure}[t!]
	\includegraphics[width=\columnwidth]{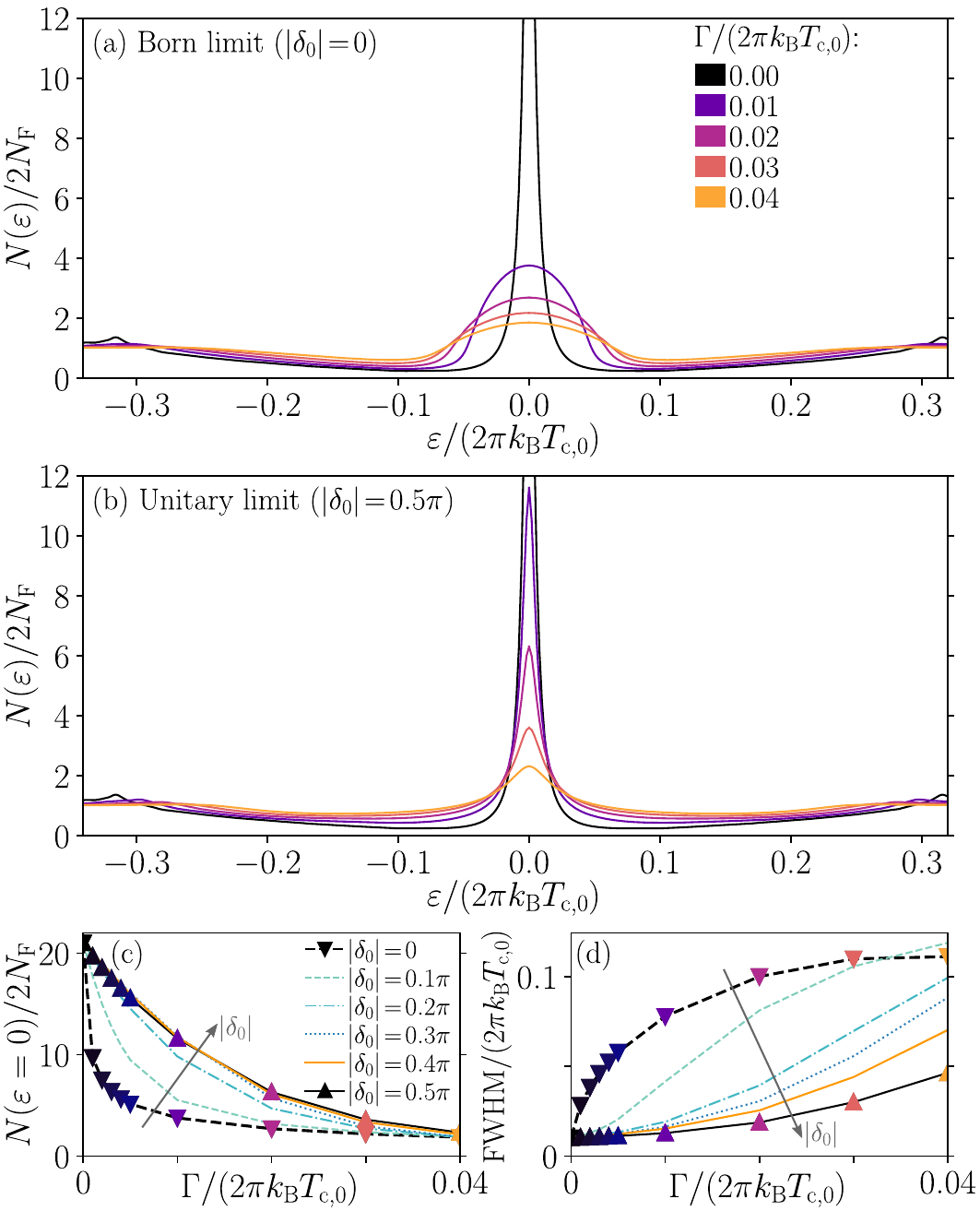}
	\caption{LDOS as a function of energy in the low-energy regime and at a maximally pair-breaking $[110]$ edge of a $d_{x^2-y^2}$-wave superconductor with the same geometry as in Fig.~\ref{fig:systemSketch}(b)
    at $T=0.5T_{\rm c}$, illustrating the broadening of midgap ABS by impurities in the (a) Born ($\delta_0=0$) and (b) unitary ($|\delta_0|=\pi/2$) scattering limits. Colors denote the scattering energy $\Gamma$. (c),(d) Peak height and full width at half maximum (FWHM), respectively, as functions of $\Gamma$ for several different scattering phase shifts $\delta_0$. The calculations were performed with a small phenomenological broadening $\eta = 0.005 \times 2\pi\kb T_{{\rm c},0}$ to ensure numerical stability. This yields a finite peak height and width even as $\Gamma \to 0$, while without broadening the ABS peak is a delta distribution~\cite{lofwander_andreev_2001}.}
	\label{fig:BroadeningABS}
\end{figure}

The phase crystal realization studied in this work relies on the presence of low-energy ABS.
In order to better understand the results in the rest of the work, we first in this section briefly explain the origin of surface ABS, before investigating their broadening by impurities in the absence of the phase crystal instability in Fig.~\ref{fig:BroadeningABS}.

Surfaces of unconventional superconductors are known to host interesting phenomena such as bound states~\cite{Sigrist1991Apr,Matsumoto1995Aug,Nagato1995Jun,Fogelstrom1997Jul,Sigrist1998Jun,Honerkamp2000May,Iniotakis2008Jan,Zare2010Jun,Potter2014Mar,Suzuki2014May,Vorontsov2018,Wilcox2022Mar}.
In particular, at a $[110]$ edge of a $d_{x^2-y^2}$-wave superconductor, quasiparticle scattering between $d$-wave lobes of opposite sign (see Fig.~\ref{fig:systemSketch}) leads effectively to an accumulation of a phase factor relative to the condensate.
This phase factor leads to resonant Andreev reflection and consequently surface pair-breaking, resulting in highly degenerate zero-energy ABS~\cite{Hu1994,Jian1994,kashiwaya_tunnelling_2000,lofwander_andreev_2001,Sato2011Jun}.
We illustrate these zero-energy ABS in Fig.~\ref{fig:BroadeningABS} by plotting the low-energy LDOS at the $[110]$ edge, where line colors correspond to different values of the scattering energy $\Gamma$ in the (a) Born scattering limit and (b) unitary scattering limit.
Here, impurity scattering connects all momentum directions such that surface scattering no longer only connects order parameter lobes of opposite sign, but also of the same sign.
Consequently, the phase factor is reduced on average, thus also reducing the density of zero-energy ABS.
We therefore observe a suppression of the ABS peak in Fig.~\ref{fig:BroadeningABS}(c) and a broadening in Fig.~\ref{fig:BroadeningABS}(d) as a function of increasing $\Gamma$, where we plot results for several different scattering phase shifts $\delta_0$.
We find that the broadening generally increases monotonically with smaller $|\delta_0|$ (see arrows), being the largest (smallest) in the Born (unitary) scattering limit, which is consistent with analytical results from earlier studies on the ABS broadening~\cite{Poenicke1999,Zare2008Sep}.
However, the broadening saturates at sufficiently large $\Gamma$ in the Born limit, which breaks the monotonic behavior and has consequences for the emergent TRSB phases (Sec.~\ref{sec:phase_diagram}).
Specifically, the energy range of $\Gamma$ % \in [0,0.04] \times 2\pi\kb T_{\mathrm{c},0}$
in Fig.~\ref{fig:BroadeningABS} is the relevant range for the phase crystal state as we show in the subsequent sections, whereas beyond this range the system becomes increasingly disordered with fully broadened states, where it thus makes less sense to talk about distributions and widths.

\section{Phase crystals and impurities}
\label{sec:phase_crystal}
Having understood the impurity effect on the low-energy ABS, we next turn to how impurities influence the phase crystal occurring due to the low-energy ABS. In order to better understand the phase crystal state itself, we start this section by briefly discussing the phase crystal instability.
Based on this discussion, we then proceed to investigate and explain how impurities influence the phase crystal.

\begin{figure*}[t!]
	\includegraphics[width=\textwidth]{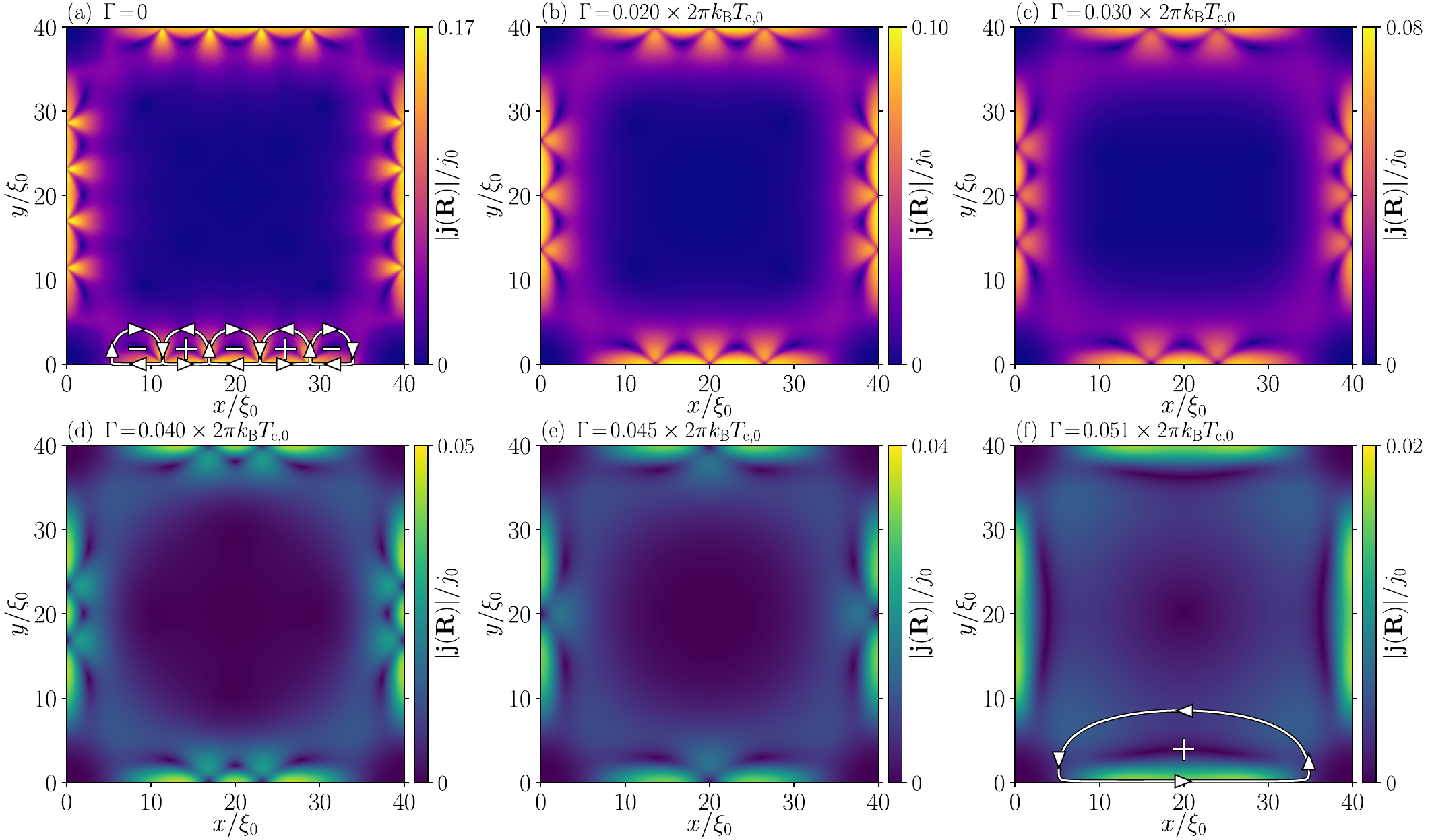}
    \caption{Magnitude of the spontaneous current density $\mathbf{j}$ as a function of coordinate $\RR$ for different scattering energies $\Gamma$, in a $d$-wave superconducting square with sidelength $\mathcal{D}=40\xi_0$ [same as Fig.~\ref{fig:systemSketch}(a)] in the unitary limit ($|\delta_0|=\pi/2$) at low temperature $T=0.02 T_{\mathrm{c},0}$. As $\Gamma$ increases from (a) to (f), there is a monotonic reduction in the current magnitude (see changing color scales) and number of current loops [see graphics in (a) and (f)]. The average number of loops per edge are (a) $5$, (b) $4$, (c) $3.5$, (d) $3$, (e) $2$ and (f) $1$. Here, (a)--(e) correspond to the phase crystal state~\cite{Holmvall2020Jan}, while (f) corresponds to the Vorontsov phase~\cite{Vorontsov2009Apr}.
    }
    \label{fig:currentSuppression_unitary}
\end{figure*}

\subsection{Phase crystal instability}
\label{sec:phase_crystal_background}
We here briefly summarize the energetics that drive the phase crystal instability and determine its emergent properties, specifically its finite periodicity and spontaneous currents.

The zero-energy ABS studied in Sec.~\ref{sec:ABS} are associated with broken superconducting pairs and thus an energy cost $\delta\Omega_{\rm ABS} \propto \Delta_0$. %, which is furthermore effectively weighted by $\Delta_0/T$ in the propagator $\hat{g}$ (Sec.~\ref{sec:theory:clean_system}).
Consequently, these states become thermodynamically unstable as $T\rightarrow 0$, and any shift $\delta_\varepsilon$ to finite energies lowers the free energy by $\delta\Omega_{\rm shift} \propto -\delta_\varepsilon$~\cite{Vorontsov2018}.
For instance, any superflow $\mathbf{p}_{\rm s}(\mathbf{R}) = (\hbar/2)\nabla\chi(\mathbf{R}) - (e/c)\mathbf{A}(\mathbf{R})$ leads to a Doppler shift $\delta_\varepsilon \propto \mathbf{v}_{\rm F}\cdot\mathbf{p}_{\rm s}$~\cite{Aprili1999Nov,Vorontsov2018}, and is furthermore associated with currents and magnetic fields (see further below).
This implies that the system becomes unstable towards spontaneous TRSB with spontaneous superflow~\cite{Higashitani1997Apr,Barash2000Sep,Lofwander2000Dec}.
The transition temperature $T^{*}$ of such an instability is essentially determined by the energy gain of the Doppler shift close to the surface, $\delta\Omega_{\rm shift} \propto -|\mathbf{v}_{\rm F}\cdot\mathbf{p}_{\rm S}|$, vs the kinetic energy cost of the condensate backflow further from the surface, $\delta\Omega_{\rm kin} \propto +|\mathbf{v}_{\rm F}\cdot\mathbf{p}_{\rm S}|^2$~\cite{Vorontsov2018}.
The phase crystal instability optimizes this trade-off by minimizing the backflow to the same short decay length $y_0\propto\xi_0$ as the surface ABS (surface normal $\hat{y}$), and is thus typically the most energetically favorable TRSB scenario with a high $T^*$ (e.g.~$T^* \approx 0.2 T_{\mathrm{c},0}$ in a clean nodal $d$-wave system~\cite{Hakansson2015Sep,Holmvall2018Jun,Holmvall2020Jan,Chakraborty2022Apr}).
This finite decay length $y_0$ is achieved through translational symmetry breaking (TSB) along the surface via the inhomogeneous ground-state solution $\chi(x,y) \propto -(1-y/y_0)e^{-y/y_0}\cos(q_xx)$~\cite{Holmvall2020Jan}, with order parameter phase $\chi$.
This solution emerges from the minimization of the kinetic energy
$F_{\rm kin} = \int d\RR d\RRprime \ps(\RR)\cdot\vDs(\RR,\RRprime)\cdot\ps(\RRprime) + \mathcal{O}(\ps^4)$,
where spontaneous TSB is related to how the nonlocal superfluid stiffness tensor $\vDs(\RR,\RRprime)$~\cite{Xu1995Jun} couples a finite superflow decay length $y_0$ to a finite wavenumber $q_x$ along the interface, via $q_x \propto y_0^{-1}$~\cite{Holmvall2020Jan}.
Additionally, we note that shorter periods $q_x^{-1} \ll \xi_0$ with rapid oscillations are penalized via the off-diagonal tensor components, and also lead to smaller superflow on average due to the frequent sign changes.
Consequently, the ground state corresponds to a phase crystal state with optimal period vs decay length~\cite{Holmvall2020Jan}.
Finally, the superflow of the phase crystal drives equilibrium current loops through the nonlocal response $\vj(\RR) = \int d\RRprime\ps(\RRprime)\cdot\vDs(\RR,\RRprime) + \mathcal{O}(\ps^2)$~\cite{Holmvall2020Jan}, see Fig.~\ref{fig:systemSketch}, which is related to a spontaneous magnetic field via Eq.~(\ref{eq:ampere_law}).
We clarify that the current loops have the same periodicity as the inhomogeneous phase $\chi(x,y)$.

\subsection{Impurity effects on phase crystals}
\label{sec:current_weakening}
Having described the phase crystal instability above we in this subsection establish how impurities influence the phase crystal, focusing in particular on the magnitude and periodicity of the spontaneous current as that gives a clear fingerprint of the overall behavior.

In a finite-sized system, the periodicity of the phase crystal gives rise to a finite number of current loops, each with size $\sim 5\xi_0$, see Fig.~\ref{fig:currentSuppression_unitary}.
Here, more loops imply on average a smaller current, and thus small superflow, since they have to change sign more often, but also with smaller backflow into the bulk as described in Sec.~\ref{sec:phase_crystal_background}.
Furthermore, the maximum current flow is found in a clean system at zero temperature, because the energy cost and density of zero-energy ABS is then the largest.
The phase crystal is thus suppressed by any broadening or reduction of the zero-energy ABS, which has already been shown to be the case for e.g.~higher temperature~\cite{Hakansson2015Sep}, external flux~\cite{Holmvall2018Jun}, geometric effects~\cite{Holmvall2019May}, Fermi surface effects~\cite{Wennerdal2020Nov}, and nonmagnetic impurities~\cite{Chakraborty2022Apr,Seja2022Oct}. 

We show in Figs.~\ref{fig:currentSuppression_unitary}(a)--\ref{fig:currentSuppression_unitary}(f) that as the scattering energy $\Gamma$ increases, there is a gradual weakening of the current density $\mathbf{j}(\RR)$, directly caused by the impurity broadening of zero-energy ABS.
This is explained by that the reduced amount of zero-energy ABS leads to less possible energy gain by Doppler shift, vs essentially the same cost of the backflow response.
We also find that increased $\Gamma$ (or reduced $T$) leads to a reduced number of current loops as depicted in Fig.~\ref{fig:currentSuppression_unitary}, and thus an increased phase crystal period.
This reduction of the number of current loops is more subtle; to directly quantify and explain it, we establish different self-consistent solutions with different number of loops and compare their free energy, as a function of both temperature and impurity scattering energy.
To summarize this involved analysis (see Appendix~\ref{app:current_loops} for further details), we find that the reduction is mainly related to two effects.
First, the impurity suppression of the $d$-wave order parameter $\Delta(\Gamma)$ leads to an increased coherence length $\xi(T)/\xi_0 \approx \Delta_0/\Delta(\Gamma)$~\cite{Abrikosov1960Dec,Poenicke1999}, such that the phase crystal has to change to longer (i.e.~fewer) current loops to maintain the same effective periodicity.
Second, the reduction in magnitude as described above can be compensated by less current loops since this changes sign less often (i.e.~an average stronger Doppler shift).
Additionally, we find that the total number of current loops generally change in multiples of two, but note that certain geometric effects can lead to a change by e.g.~single loops at a time~\cite{Holmvall2018Mar}.
This is related to the minimization of net current and net orbital magnetization, which both cancel to zero for an equal number of loops with negative and positive circulation direction (thus an even number of loops in total).
However, they can also cancel on average for odd number of loops if the loop sizes differ~\cite{Holmvall2018Mar,Holmvall2019May}.

While the above behavior of reduced magnitude and number of current loops occurs for both the square and triangle-like geometries, there are unique finite-size effects in a mesoscopic system such as the square.
In particular, ABS at different pair-breaking edges may hybridize in the mesoscopic square, while the triangular system hosts no such hybridization due to having only a single pair-breaking edge, see Fig.~\ref{fig:systemSketch}.
Here, the hybridization in the square becomes stronger with higher scattering energy since it increases the effective coherence length $\xi(\Gamma)$, thus reducing the effective system size $\mathcal{D}/\xi(\Gamma)$ (we also verify this by changing $\mathcal{D}$ at fixed $\Gamma$).
%Here, $\xi_0$ ($\Delta_0$) is the bulk coherence length (gap) at $T=\Gamma=0$.
Thus, as $\mathcal{D}/\xi(\Gamma)$ decreases we find that the hybridization can eventually become so strong that there is effectively no bulk superconductivity established in the square system~\cite{Holmvall2019May}, consequently changing the energy balancing such that the phase crystal period vs decay length is eventually no longer relevant.
At this point, the phase crystal competes with another TRSB phase with spatially uniform superflow and currents along each pair-breaking edge, see Fig.~\ref{fig:currentSuppression_unitary}(f).
This latter phase was originally proposed by Vorontsov~\cite{Vorontsov2009Apr} and is
triggered by the hybridization between different pair-breaking edges, and has previously been studied mainly in a slab geometry~\cite{Hachiya2013Aug,Higashitani2015Jan,Miyawaki2015Mar,Miyawaki2015Nov,Miyawaki2017Oct,Miyawaki2018Oct}.
Our results therefore highlight that it is possible for the Vorontsov phase to emerge in a fully confined system such as the mesoscopic square, due to impurities reducing $\mathcal{D}/\xi(T)$ (see also Ref.~\cite{Hakansson2015Sep} for a clean system).
In contrast, we find that this phase does not emerge in the triangle-like geometry since it has only a single pair-breaking edge.

Finally, we note that a generalized Poincaré-Hopf theorem relates the number topological defects in the superfluid momentum $\mathbf{p}_\mathrm{s}(\RR)$ (and thus number of current loops) to the Euler characteristic of the superconductor~\cite{Holmvall2018Jun}, such that a variation of the number of loops or geometry leads to compensating topological defects in $\mathbf{p}_\mathrm{s}(\RR)$.
Thus, our results here imply that tuning the impurity concentration can be used to directly control the topological defect structure in $\mathbf{p}_\mathrm{s}(\RR)$.

\section{Ground-state phase diagram}
\label{sec:phase_diagram}
In Sec.~\ref{sec:current_weakening} we established how impurities reduce the magnitude and periodicity of the phase crystal currents, increase mesoscopic finite-size effects, and induce the competing Vorontsov phase~\cite{Vorontsov2009Apr} which instead hosts homogeneous currents along the edges.
To fully quantify these effects, we finally here present the full ground-state phase diagram as a function of temperature $T$ and impurity scattering energy $\Gamma$, for different scattering phase shifts $\delta_0$.
We do this for both the square and triangle-like geometries shown in Fig.~\ref{fig:systemSketch}, i.e.~with and without significant mesoscopic finite-size effects acting on the ABS, respectively.
We begin by focusing on the transition temperature $\tc(\Gamma)$ between the normal state into a superconducting state with preserved translational symmetry and time-reversal symmetry (TRS), i.e.~without phase crystals.
We then focus on the transition temperature $T^*(\Gamma)$ into the TRSB phases.
Finally, we investigate the competition between the phase crystal and the Vorontsov phase.

\begin{figure}[t!]
	\includegraphics[width=\columnwidth]{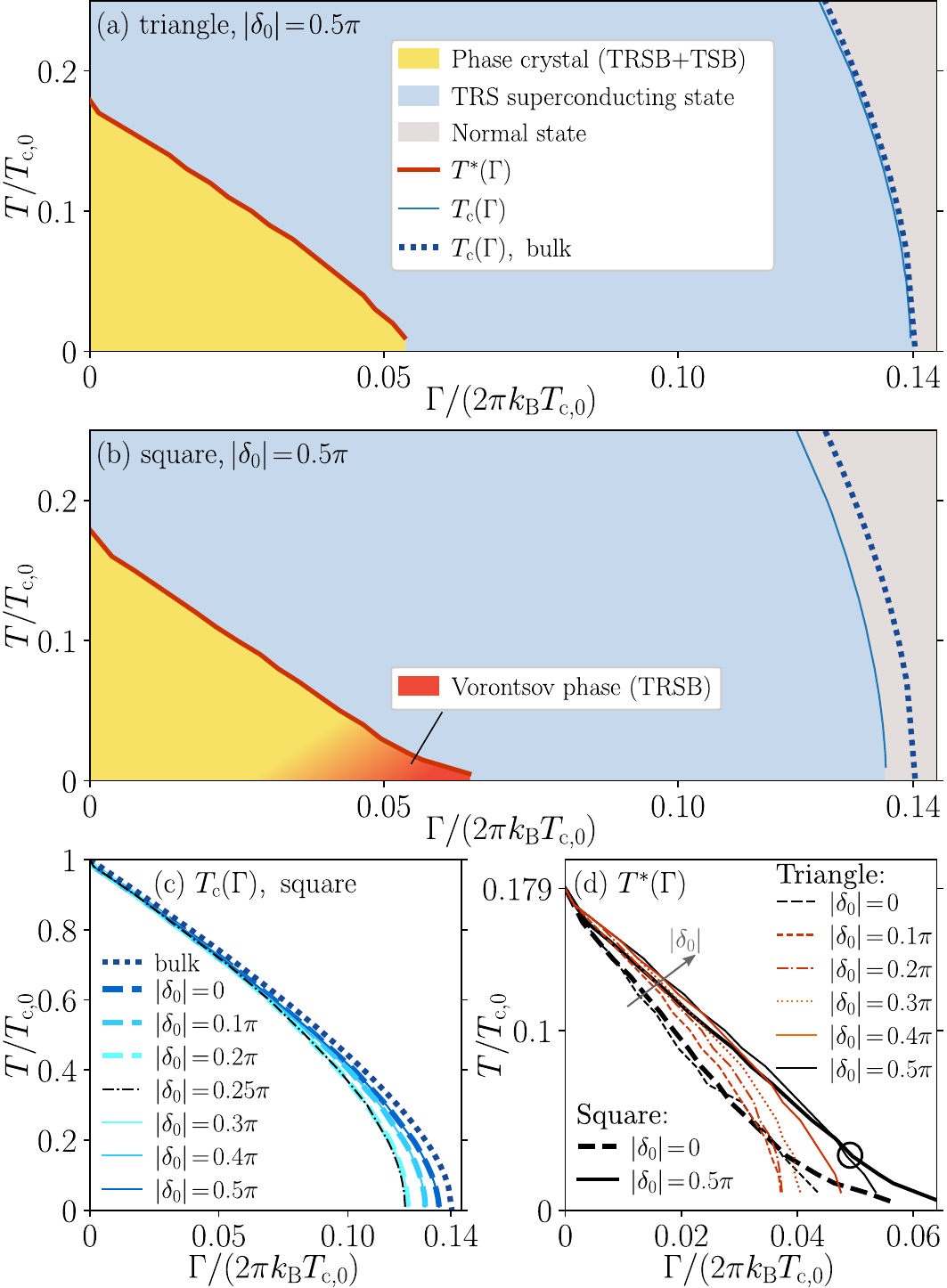}
    \caption{(a)-(b) Ground-state phase diagram as a function of temperature $T$ and impurity scattering energy $\Gamma$ in the unitary scattering limit ($|\delta_0|=0.5\pi$), for the triangle and square geometries in Fig.~\ref{fig:systemSketch}, respectively. Here, $\tc(\Gamma)$ is the transition temperature between the normal state (gray) and the superconducting state with preserved time-reversal and translational symmetries (blue) in the finite-sized system (thin solid line) and in the Abrikosov-Gor'kov bulk limit (thick dashes). $T^*(\Gamma)$ (thick solid line) is the transition between the superconducting state and the phase crystal (yellow) or the Vorontsov phase~\cite{Vorontsov2009Apr} (red). (c) Comparison of $\tc(\Gamma)$ in the square geometry for different scattering phase shifts $\delta_0$. (d) Comparison of $T^*(\Gamma)$ in the triangle and square geometries for different $\delta_0$. Here, the circle indicates relative enhancement of $T^*(\Gamma)$ at large $\Gamma$ in the square compared to the triangle system due to the onset of the Vorontsov phase, which is absent in the triangle geometry. As a reference for the energy scales, the clean bulk $d$-wave order parameter at $T=0$ is $\Delta_0/(2\pi\kb T_{\mathrm{c},0}) \approx 0.24$.
    }
    \label{fig:phase_diagram}
\end{figure}

Figures~\ref{fig:phase_diagram}(a) and \ref{fig:phase_diagram}(b) show the ground-state phase diagram in the triangle-like and square geometries, respectively, as a function of temperature $T$ and impurity scattering energy $\Gamma$ in the unitary scattering limit (see further below for other $\delta_0$).
We first discuss the triangle-like system followed by the square system.
In Fig.~\ref{fig:phase_diagram}(a), we find an excellent agreement in $\tc(\Gamma)$ between the triangular system and the bulk Abrikosov-Gor'kov result, Eq.~(\ref{eq:AbrikosovGorkovFormula}). This
$\tc(\Gamma)$ thus shows a bulk-like behavior, which we find for other $\delta_0$ and relate to the minimal influence of the single pair-breaking edge at higher temperatures.
This can be understood by studying the propagator $\hat{g}$ in Eqs.~(\ref{eq:GreensFunctionSurface:1})--(\ref{eq:GreensFunctionSurface:2}) in Sec.~\ref{sec:theory:clean_system}, which is bulk-like at all edges except at the single pair-breaking edge where there is an additional bound-state term but which is significantly suppressed at elevated temperatures.
In contrast, all edges are pair-breaking in the square-shaped system in Fig.~\ref{fig:phase_diagram}(b), such that there is finite edge-edge hybridization between ABS at different edges and a faster suppression of the order parameter~\cite{Vorontsov2009Apr}.
Consequently, we find a stronger suppression of $\tc(\Gamma)$ in the square-shaped system due to these mesoscopic finite-size effects, which we investigate in greater detail for different scattering phase shifts $\delta_0$ in Fig.~\ref{fig:phase_diagram}(c).
Here, the edge-edge hybridization increases with $\Gamma$ since this effectively reduces the system size $\mathcal{D}/\xi(\Gamma)$, eventually leading to a loss of bulk superconductivity at the system center, thus with subgap states in the entire system~\cite{Holmvall2019May}.
Furthermore, since the broadening of these subgap states depends on the scattering phase shift $\delta_0$ (see Sec.~\ref{sec:ABS}), the hybridization and suppression of $\tc(\Gamma)$ therefore also depend on $\delta_0$, as clearly seen in Fig.~\ref{fig:phase_diagram}(c).
Interestingly, we find that $\tc(\Gamma)$ has a local minimum approaching $|\delta_0| = \pi/4$ and is symmetric around this point, e.g.~the Born and unitary limits have the same $\tc(\Gamma)$.
Additionally, the impurity suppression of $\tc(\Gamma)$ naturally becomes more pronounced with smaller system size $\mathcal{D}$ as described above, such that $\tc(\Gamma)$ tends towards zero for decreasing $\mathcal{D}$.
In the opposite limit of increasing $\mathcal{D}$, the results of the square system also approach the bulk limit as in the triangle-like system.

Next, we focus on the transition temperature $T^*(\Gamma)$ into the TRSB states, which depends on the scattering phase shift $\delta_0$ in both the triangle-like and square geometries, see Fig.~\ref{fig:phase_diagram}(d).
This stems from how $\delta_0$ influences the ABS energy broadening at a given $\Gamma$ [Fig.~\ref{fig:BroadeningABS}(d)], which in turn influences the energy gain by the superflow-mediated Doppler shifts.
Specifically, comparing the Born (black dashed) and unitary (black solid) scattering limits, the transition temperature starts at $T^*(\Gamma=0)\approx0.179 T_{\mathrm{c},0}$ in the clean limit~\cite{Hakansson2015Sep,Holmvall2018Jun}, with a roughly linear $T^*(\Gamma)$ over the entire phase diagram in both systems.
However, the phase crystal is more robust in the unitary scattering limit due to a smaller broadening of zero-energy ABS compared to the Born limit as established in Sec.~\ref{sec:ABS}, such that there is more energy to gain by the Doppler shifts.
Importantly, in terms of the critical impurity strength $\Gamma^*(T)$ we find that at low temperatures the phase crystal survives up until roughly $50\%$ ($40\%$) of the superconducting critical impurity strength %$\Gamma_{\rm c}$
in the unitary (Born) scattering limit.
Results for other scattering phase shifts (colors) fall monotonically between these two limits, see arrow in Fig.~\ref{fig:phase_diagram}(d), which directly follows how the ABS broadening changes monotonically with $|\delta_0|$ [see Fig.~\ref{fig:BroadeningABS}(d)].
The only exception to this behavior is the saturation in ABS broadening in the Born limit at large $\Gamma$ as discussed in Sec.~\ref{sec:ABS}, i.e.~see the kink in the thin dashed black line at $\Gamma\approx 0.03\times 2\pi\kb T_{\mathrm{c},0}$ in Fig.~\ref{fig:phase_diagram}(d).
Regardless, our results demonstrate a disorder-robust phase crystal in weak-coupling superconductors, for all scattering phase shifts.

Finally, we note that there is minimal difference in $T^*(\Gamma)$ between the square and triangular systems except at large $\Gamma$, where the TRSB state survives for larger values in the square, see the deviation of the black curves at small $T$ in Fig.~\ref{fig:phase_diagram}(d) marked by a circle for $|\delta_0|=\pi/2$ (no circle is shown for $\delta_0=0$ because of visibility reasons).
We find that this enhancement of $T^*(\Gamma)$ in the square system relative to the triangle is associated with the emergence of the phase proposed by Vorontsov~\cite{Vorontsov2009Apr}, i.e.~induced by mesoscopic finite-size effects present in the square system [see Fig.~\ref{fig:currentSuppression_unitary}(f)] but completely absent in the triangle-like system.
Furthermore, we find that the Vorontsov phase only appears for smaller $T$ and larger $\Gamma$, e.g. in the unitary limit it appears below $T \approx 0.02 T_{\mathrm{c},0}$ and above $\Gamma \approx 0.05 \times 2\pi\kb T_{\mathrm{c},0}$ for the system size $\mathcal{D}=40\xi_0$ studied here, see Fig.~\ref{fig:phase_diagram}(b).
%, while in the Born limit it appears below $T \approx 0.01 T_{\mathrm{c},0}$ and above $\Gamma \approx 0.04 \times 2\pi\kb T_{\mathrm{c},0}$.
Here, the transition between the phase crystal and the Vorontsov phase is not distinct~\cite{Hakansson2015Sep}, as indicated by the color gradient.
Specifically, the yellow region hosts the periodic phase crystal, the dark orange region is the Vorontsov phase with uniform currents, while in the gradient color region in-between both states can coexist along different portions of the system edge due to having very similar free energy, and are therefore difficult to numerically separate, see also discussion in Sec.~\ref{sec:phase_crystal_background}.
We find that the Vorontsov phase covers a larger portion of the phase diagram at smaller system sizes $\mathcal{D}$~\cite{Hakansson2015Sep,Vorontsov2009Apr} as fully explained by the effective shrinking of the system size $\mathcal{D}/\xi(\Gamma)$ and consequent enhancement of the finite-size effects.

\section{Summary and outlook}
\label{sec:summary}
Phase crystals~\cite{Hakansson2015Sep,Holmvall2018Jun,Holmvall2020Jan} are a class of inhomogeneous superconducting ground states characterized by spontaneous phase gradients, which nonlocally drive inhomogeneous currents and magnetic fields, distinctly different from e.g.~Abrikosov vortices and the Fulde-Ferrell-Larkin-Ovchinnikov states.
The phase crystal instability is generally induced by an inhomogeneous and negative superfluid stiffness~\cite{Holmvall2020Jan}, which can occur e.g.~due to flat band ABS at zero energy along edges of nodal $d$-wave superconductors.
In order to aid experimental efforts to realize and detect phase crystals, we explicitly quantify their robustness against nonmagnetic impurities, omnipresent in all materials.
Specifically, we establish the ground-state phase diagram as a function of temperature $T$ and impurity strength $\Gamma$, for general scattering phase shifts $\delta_0$ between the Born and unitary scattering limits.
This is done by employing the $t$-matrix approach within the quasiclassical theory of superconductivity, with full self-consistency simultaneously in the impurity self-energies $h_{\rm imp}$, superconducting order parameter $\Delta$, and magnetic vector potential $\vA$.

We demonstrate that impurities cause an energy broadening of the flat band zero-energy ABS appearing at pair-breaking edges of nodal $d$-wave superconductors, which is consistent with early analytic calculations~\cite{Poenicke1999,Zare2008Sep}.
Specifically, we show that the broadening generally increases with $\Gamma$ and decreases with $\delta_0$.
The phase crystal instability relies on the energy gain by Doppler shifting zero-energy ABS to finite energies, vs the energy cost of the condensate backflow away from the surface.
Here, the backflow decay length is nonlocally coupled to the phase crystal periodicity along the surface via the superfluid stiffness tensor \cite{Holmvall2020Jan}.
Consequently, we find that the ABS broadening reduces the possible energy gained by forming the phase crystal, which subsequently reduces the magnitude, periodicity, and transition temperature $T^*(\Gamma)$ of the phase crystal.
However, in terms of the critical impurity strength $\Gamma^*(T)$, we find that the phase crystal is remarkably robust, surviving up until $\sim40\text{--}50\%$ of the superconducting critical impurity strength from the Born to unitary scattering limit, respectively.
Our results therefore imply that it should be possible to observe phase crystals in the presence of impurities also in weak-coupling superconductors, not just for strong electron-electron correlations as previously established~\cite{Chakraborty2022Apr,Seja2022Oct}.

We further show that in mesoscopic systems, hybridization between different pair-breaking edges can become detrimental enough to altogether suppress the superconducting transition temperature $T_{\rm c}(\Gamma)$, while it instead enhances the TRSB transition temperature $T^*(\Gamma)$.
This enhancement of $T^*(\Gamma)$ is related to a competition with another spontaneous TRSB phase proposed by Vorontsov~\cite{Vorontsov2009Apr}, which is directly induced by the edge-edge hybridization, and is characterized by translationally invariant superflow and currents along each edge.
We find that the transition between the phase crystal instability and the Vorontsov phase is smooth, and that the competition occurs mainly for large impurity scattering energies since this enhances the edge-edge hybridization by reducing effective system size $\mathcal{D}/\xi(\Gamma)$, where $\xi(\Gamma)$ is the coherence length.
In other words, reducing the system size $\mathcal{D}$ or increasing $\Gamma$ increases the competition.

Many open questions still remain regarding phase crystals and pose as interesting topics for further studies.
The ``holy grail'' is experimentally observing phase crystals. Direct detection has been proposed~\cite{Hakansson2015Sep,Holmvall2018Jun,Holmvall2019May,Chakraborty2022Apr} based on e.g.~scanning tunneling spectroscopy, nano scanning quantum interference devices, and nitrogen-vacancy centers.
Indirect detection has been proposed based on tunneling experiments and zero-bias conductance measurements~\cite{Hakansson2015Sep,Chakraborty2022Apr}, as well as the associated jump in heat capacity~\cite{Holmvall2018Jun,Holmvall2019May}, magnetic moment~\cite{Holmvall2023Mar}, and orbital magnetization~\cite{Holmvall2018Jun}.
Our work aids such experimental efforts by explicitly quantifying the robustness against nonmagnetic impurities, via the full ground-state phase diagram as a function of temperature and impurity strength, with and without mesoscopic finite-size effects, and for general scattering phase shifts.
To further establish the realizability of phase crystals in experimental systems, additional studies of robustness and observability are still important.
For instance, while we compute the phase diagram with full self-consistency in the impurity self-energies, we assume the noncrossing approximation for the $t$-matrix.
The influence of such contributions together with full self-consistent impurity energies is therefore an interesting future outlook.
Furthermore,
% Yamada1996March
Nagato et al.~\cite{Nagato1996Apr,Yamada1996March,Nagato1998Mar,Higashitani2024Dec} have used generalized boundary conditions for surface roughness to study the influence on other edge modes and symmetry-broken phases, while Suzuki et al.~\cite{Suzuki2015Jun,Suzuki2016Oct,Suzuki2024Jun} have considered a local impurity concentration.
Our results imply that phase crystals should be stable against such surface roughness but only until some critical roughness where the broadening of surface ABS is too large.
Quantifying this transition becomes important to understand the robustness of phase crystals.
Similarly, interface transparency towards other materials leads to surface ABS broadening which is thus expected to weaken the phase crystal~\cite{Holmvall2019Nov}, but also poses as an interesting approach for experimental measurement by tunneling through the phase crystal state.
We note that spontaneous current loops have also been studied in the context of $s+id$ or $d+is$ superconductivity~\cite{Li2021April,Breio2022Jan,Breio2024Jan,Andersen2024May,Pathak2024Jul,Pupim2024Nov}, chiral $p$-wave superconductors~\cite{Bouhon2014Dec}, and topological superconducting nanowires~\cite{Li2013Aug,Wang2014May}.
An interesting outlook is therefore to study these states in a similar approach as the phase crystal, namely via the nonlocal superfluid stiffness tensor and the superflow~\cite{Holmvall2020Jan}.
Other highly interesting questions involve identifying entirely different systems where phase crystals may appear.

%%%%%%%%%%%%%%%%%%%%%%%%%%%%%%%%%%%%%%%%%%%
\acknowledgements
%%%%%%%%%%%%%%%%%%%%%%%%%%%%%%%%%%%%%%%%%%%
We acknowledge M.~H\r{a}kansson, O.~Shevtsov, and P.~Stadler for their work on SuperConga.
KMS, NWW, TL, and MF were partially supported by Vetenskapsr{\r{a}}det (VR) grant 2020-05261 and Chalmers Excellence Initiative Nano.
AMBS and PH were partially supported by the Knut and Alice Wallenberg Foundation, KAW 2019.0309, through the Wallenberg Academy Fellows program, and the European Union through the European Research Council (ERC) under the European Union’s Horizon 2020 research and innovation programme (ERC-2022-CoG, Grant
agreement No. 101087096). Views and opinions expressed are however those of the authors only and do not necessarily reflect those of the European Union or the European Research Council Executive Agency. Neither the European Union nor the granting authority can be held responsible for them.
The computations and data handling were enabled by resources provided by the National Academic Infrastructure for Supercomputing in Sweden (NAISS) at C3SE, NSC, PDC and HPC2N, partially funded by the Swedish Research Council through grant agreements No.~2022-06725.
Additional computations were enabled by the Berzelius resource provided by the Knut and Alice Wallenberg Foundation at the National Supercomputer Centre.
%\section*{Data availability}
All data is publicly available~\cite{Seja2024Dec:Data} and was generated using the open-source framework \textsc{SuperConga}~\cite{Holmvall2023Mar}.

% Appendix
\appendix

\section{Free energy analysis of phase crystal periodicity}
\label{app:current_loops}
In this Appendix we analyze how the temperature $T$ and impurity scattering energy $\Gamma$ influence the number of current loops $n$ of the phase crystal, thus complementing the discussion in Sec.~\ref{sec:current_weakening}.
Specifically, we compare the free energy $\Omega$ [Eq.~(\ref{eq:FreeEnergyEilenberger})] as a function of $T$ and $\Gamma$ for different self-consistent solutions with different $n$.
In general, we find that these energy differences are quite small compared to other energy scales (e.g.~$\Delta_0$), but still large enough to make some configurations metastable or even unstable.
We note that the smallness of energy comes partly from that we normalize by system size, such that the phase crystal energy here is suppressed by the edge-to-area ratio.

\begin{figure}[t!]
	\includegraphics[width=\columnwidth]{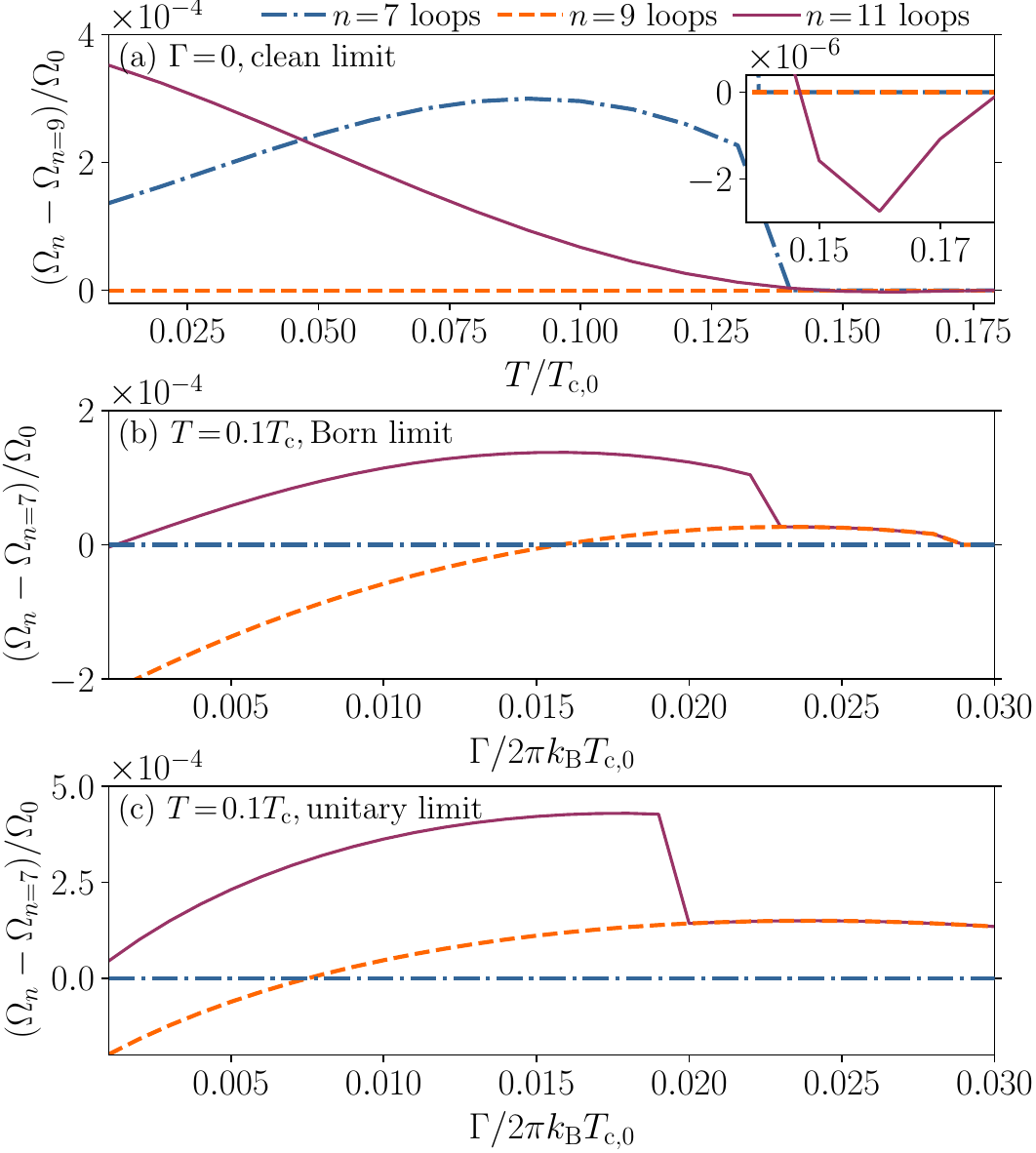}
    \caption{(a) Difference in free energy $\Omega$ between phase crystals with different number of current loops $n$ in the triangle-like system [Fig.~\ref{fig:systemSketch}(b)], as a function of temperature $T$ at fixed impurity scattering energy $\Gamma=0$ (i.e.~the clean limit). The inset is a zoom of the same figure. (b),(c) Same as (a) but as a function of $\Gamma$ at fixed $T=0.1\tc$ in the Born and unitary limits, respectively. Here, we use natural units $\Omega_0 \equiv \mathcal{A}\NF(\kb T_{\mathrm{c},0})^2$. Straight diagonal lines that merge two solutions correspond to a phase transition into the solution with the lowest free energy.}
    \label{fig:freeEnergyVsLoops}
\end{figure}

Figure~\ref{fig:freeEnergyVsLoops}(a) shows the free energy difference between configurations with different number of current loops $n=11,9,7$, as a function of $T$ in the clean limit ($\Gamma=0$).
Per definition, the ground state solution has the lowest free energy, while all other solutions are either metastable or unstable.
Specifically, at certain parameter ranges the metastable solutions become unstable, such that the system spontaneously transitions to a more stable solution, see for instance $n=7$ becoming unstable for $T>0.14 T_{\mathrm{c},0}$.
In general, we find that when decreasing the temperature at fixed impurity strength, it is energetically favorable to decrease the number of current loops, specifically from $n=11$ to $n=9$ loops at $T\approx 0.15 T_{\mathrm{c},0}$, see inset in \ref{fig:freeEnergyVsLoops}(a).
The reason is that the reduced number of loops leads to less sign changes in both the current and the superflow, thus with larger Doppler shifts which counteract the increased energy cost of zero-energy ABS at lower temperatures (see also discussion in Sec.~\ref{sec:current_weakening}).

Figures~\ref{fig:freeEnergyVsLoops}(b) and \ref{fig:freeEnergyVsLoops}(c) show the same as \ref{fig:freeEnergyVsLoops}(a) but as a function of $\Gamma$ at fixed $T=0.1 T_{\mathrm{c},0}$ in the Born and unitary limits, respectively.
Similar to above, we find that varying $\Gamma$ changes the ground state solution and can also make certain solutions unstable.
We find that as $\Gamma$ increases, it becomes energetically favorable with fewer current loops which we explain by mainly two effects.
First, the increased influence of impurities leads to an effectively larger coherence length and thus an effectively smaller system size, such that the system has to change to less current loops to maintain the same effective periodicity.
Second, the broadening of ABS (Sec.~\ref{sec:ABS}) and consequent reduction in current magnitude (Sec.~\ref{sec:current_weakening}) means that maintaining the same level of Doppler shift has to be achieved by more homogeneous superflow and thus less current loops.
We find that these effects are more pronounced for the unitary scattering limit compared to the Born scattering limit, i.e.~the system transitions to fewer loops at a lower impurity scattering energy for the unitary limit.
Finally, we note that when varying both $T$ and $\Gamma$, there are some less trivial effects that are for instance related to the slightly nonlinear behavior in the phase diagram in Fig.~\ref{fig:phase_diagram} in Sec.~\ref{sec:phase_diagram}.

\bibliography{references}

\end{document}